\documentclass[aps,prb,superscriptaddress,reprint,floats,citeautoscript,floatfix,nobibnotes,nofootinbib]{revtex4-1}

\usepackage[intlimits,tbtags]{amsmath}
\usepackage{color}
\usepackage{amssymb}
\usepackage{graphicx}
\usepackage{bm}
\usepackage{array}
\usepackage{dcolumn}
\usepackage{bm}

\newcolumntype{d}[1]{D{.}{.}{#1} }

\newcommand{\jsnu}{Laboratory of Quantum Materials Design and Application, School of Physics and Electronic Engineering, Jiangsu Normal University, Xuzhou 221116, China}

\newcommand{\zhanggroup}{State Key Laboratory of Superhard Materials, Key Laboratory of Automobile Materials of MOE, and College of Materials Science and Engineering, Jilin University, Changchun 130012, China}

\begin{document}
\title{Helium Incorporation Stabilized Direct-gap Silicides}

\author{Shicong Ding}
\affiliation{\jsnu}
\author{Jingming Shi}\email{jingmingshi@jsnu.edu.cn}
\affiliation{\jsnu}
\author{Jiahao Xie}
\affiliation{\zhanggroup}
\author{Wenwen Cui}
\affiliation{\jsnu}
\author{Pan Zhang}
\affiliation{\jsnu}
\author{Kang Yang}
\affiliation{\jsnu}
\author{Jian Hao}
\affiliation{\jsnu}
\author{Meiling Xu}
\affiliation{\jsnu}
\author{Qingxin Zeng}
\affiliation{\jsnu}
\author{Lijun Zhang}
\affiliation{\zhanggroup}
\author{Yinwei Li}\email{yinwei\_li@jsnu.edu.cn}
\affiliation{\jsnu}

\date{\today}

\begin{abstract}


The search of direct-gap Si-based semiconductors is of great interest due to the potential application in many technologically relevant fields. This work examines the incorporation of He as a possible route to form a direct band gap in Si. Structure predictions and first-principles calculations have shown that He reacts with Si at high pressure, to form the stable compounds Si$_2$He and Si$_3$He. Both compounds have host-guest structures consisting of a channel-like Si host framework filled with He guest atoms. The Si frameworks in two compounds could be persisted to ambient pressure after removal of He, forming two pure Si allotropes. Both Si--He compounds and both Si allotropes exhibit direct or quasi-direct band gaps of 0.84--1.34 eV, close to the optimal value ($\sim$1.3 eV) for solar cell applications. Analysis shows that Si$_2$He with an electric-dipole-transition allowed band gap possesses higher absorption capacity than diamond cubic Si, which makes it to be a promising candidate material for thin-film solar cell.

\end{abstract}
\maketitle


Pollution-free renewable energy is urgently needed as a substitute for fossil fuels. Inexhaustible solar energy is widely used, and its conversion to electricity for daily use requires photovoltaic materials.~\cite{kazmerski2006solar,jelle2012building,kannan2016solar,zhang2012genomic} Cubic diamond silicon (CD-Si) is a good candidate photovoltaic material due to its suitable band gap and stability. A good photovoltaic material should possess an electric-dipole-transition allowed direct band gap.~\cite{PhysRevLett.100.167402} The Shockley-Queisser limit~\cite{shockley1961detailed} predicts that a band gap of 1.34 eV achieves the highest solar conversion efficiency (33.7\%). However, CD-Si is an indirect-gap (1.17 eV) semiconductor, and thus not ideal for thin-film photovoltaic devices.~\cite{PhysRevLett.55.1418,PhysRevB.36.4821} Therefore, the search for new Si allotropes or Si-based compounds with an electric-dipole-transition allowed direct band gap is of great interest.

Much effort has been devoted to the search for new Si allotropes with direct or quasi-direct band gaps~\cite{wentorf1963two,ZHAO1986679,PhysRevB.50.13043,rapp2015experimental,kurakevych2016synthesis,PhysRevLett.118.146601,PhysRevLett.122.105701,huang2019synthesis,wentorf1963two,PhysRevLett.118.146601,huang2019synthesis,wang2014direct,PhysRevLett.110.118702,PhysRevB.90.115209,PhysRevB.91.214104,PhysRevB.81.115201,wu2011density,he2016direct,PhysRevLett.121.175701,wei2019six,oreshonkov2020new,he2016direct,PhysRevLett.121.175701,wei2019six,PhysRevB.86.121204,PhysRevB.92.014101,PhysRevB.62.R7707,kim2015synthesis}. A series of new Si structures formed by phase transformations under high pressure have been observed experimentally.~\cite{wentorf1963two,ZHAO1986679,PhysRevB.50.13043,rapp2015experimental,kurakevych2016synthesis,PhysRevLett.118.146601,PhysRevLett.122.105701,huang2019synthesis} In particular, direct-gap BC8-Si was formed after releasing the pressure from the high-pressure $\beta$-Sn phase to 2 GPa.~\cite{wentorf1963two} However, the relatively narrow direct band gap of 30 meV precludes BC8-Si as a photovoltaic material.~\cite{PhysRevLett.118.146601} Irradiation of amorphous Si film with a coherent electron beam stabilized a new Si$_{9}$ phase with a direct band gap of approximately 1.59 eV, indicating a potentially useful photovoltaic material.~\cite{huang2019synthesis}

First-principle calculations are important in the search for new Si structures. Structure searches based on Crystal structure AnaLYSis by Particle Swarm Optimization (CALYPSO) have found four channel-like Si allotropes (oF16-Si, tP16-Si, mC12-Si, and tI16-Si) with direct band gaps of 0.81--1.25 eV.~\cite{wang2014direct} A cubic Si$_{20}$-T phase with a quasi-direct band gap of 1.55 eV was designed using a new inverse-band-structure design approach based on CALYPSO.~\cite{PhysRevLett.110.118702} Conformational space annealing calculations have uncovered two new Si allotropes, Q135 and D135, with direct band gaps of 0.98 and 1.33 eV, respectively, both of which were proposed to be good photovoltaic materials with estimated photovoltaic efficiency of $\sim$30\%.~\cite{PhysRevB.90.115209} $Ab$ $initio$ random structure searching has also revealed a new Si structure with space group $Pbam$ and a direct band gap of 1.4 eV.~\cite{PhysRevB.91.214104} By substituting C or Ge atoms in their structures with Si atoms, at least 17 candidate structures were predicted,~\cite{PhysRevB.81.115201,wu2011density,he2016direct,PhysRevLett.121.175701,wei2019six,oreshonkov2020new} of which nine~\cite{he2016direct,PhysRevLett.121.175701,wei2019six} (M585, $Pbam$-32, $P6/mmm$, $Im\bar3m$, $C2/c$, $I4/mcm$, $I4/mmm$, $P2_1/m$, and $P4/mbm$) have direct band gaps of 0.65--1.51 eV. $Ab$ $initio$ minima hopping structure predictions have also predicted more than 44 Si structures, of which eleven ($R\bar3m$-1, $R\bar3m$-2, $C2/m$, $Immm$-1, $Immm$-2, $Immm$-3, $Pmma$, $I4_1md$, $Pnma$, $I\bar42d$ and $I2_12_12_1$) exhibit direct band gaps of 1.0--1.8 eV.~\cite{PhysRevB.86.121204,PhysRevB.92.014101} All these direct or quasi-direct Si structures are metastable, possessing a high energy relative to CD-Si, and thus are difficult to synthesize directly.

Si-rich compounds with open-framework structures formed at high pressures are good precursors to obtain new Si allotropes. A two-step synthesis method has made two metastable allotropes (a clathrate Si$_{136}$~\cite{PhysRevB.62.R7707} and a channel-like Si$_{24}$~\cite{kim2015synthesis}) by removing Na from high-pressure Na--Si compounds. Channel-like Si$_{24}$ was prepared by first synthesizing at high pressure a Na$_4$Si$_{24}$ precursor that contained a channel-like $sp^3$ Si host structure filled with linear Na chains. Na atoms were removed along the open channels via thermal degassing, leaving the pure Si$_{24}$ allotrope. Electrical conductivity and optical absorption measurements confirmed a quasi-direct band gap of 1.3 eV, making Si$_{24}$ a potential photovoltaic material.

\begin{figure}[htp]
\centering
  \includegraphics[width=0.85\linewidth,angle=0]{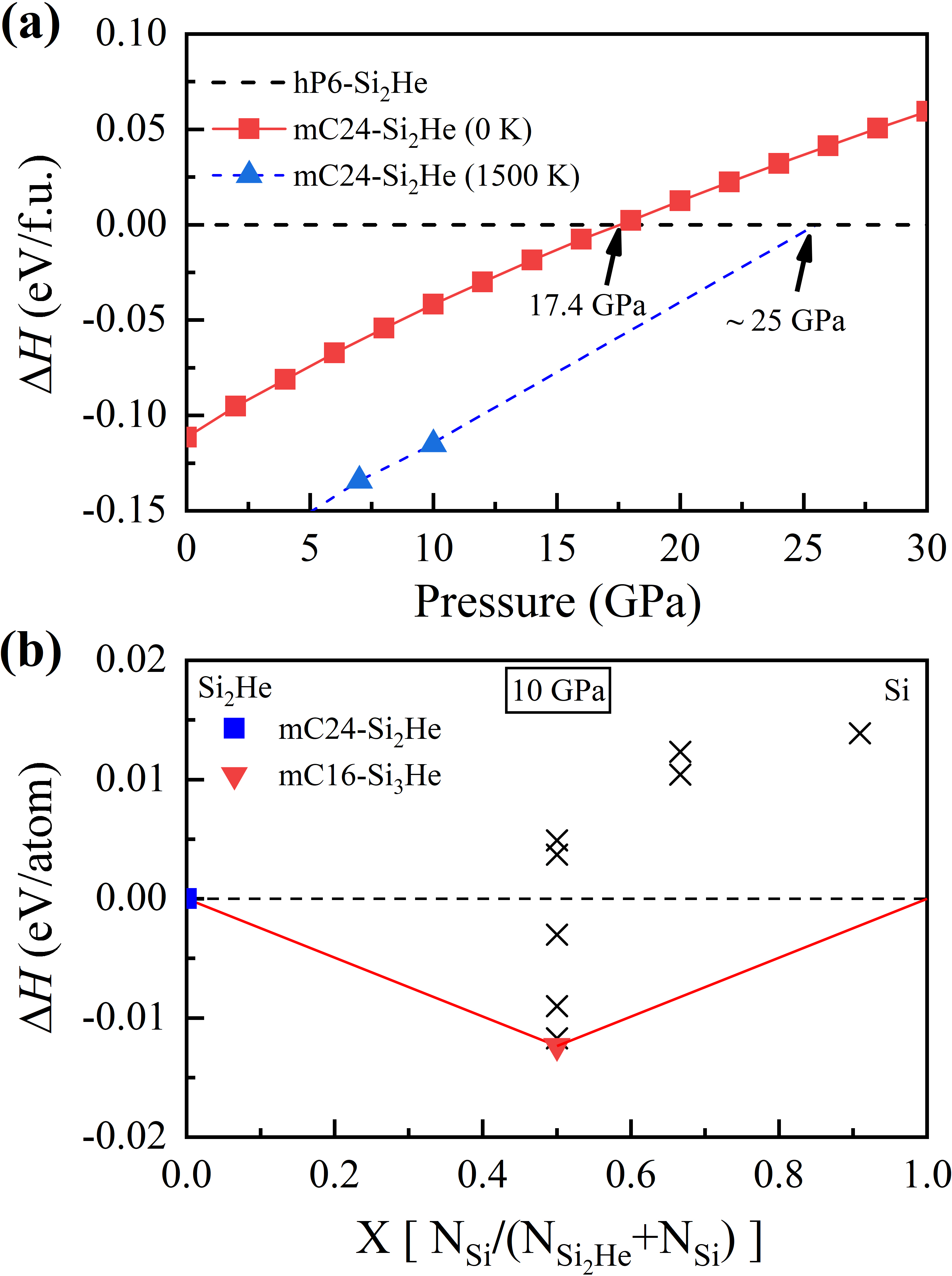}
	\caption{\label{fig:1} \textbf{(a)} Enthalpy of mC24-Si$_2$He relative to previously proposed hP6-Si$_2$He~\cite{bi2018formation} as a function of pressure at 0 and 1500 K. Arrows represent the phase transition pressures.  \textbf{(b)} Formation enthalpy of Si$_x$He$_y$ ($x$ = 1--12 and $y$ = 1--4) with respect to mC24-Si$_2$He and CD-Si at 10 GPa, defined as $\Delta H$ = [$H$(He$_xSi_y$) -- $xH$(Si$_2$He) -- ($y$ -- 2$x$)$H$(Si)] / ($x$ + $y$). Crosses represent energetically unstable structures. Compounds with formation enthalpies higher than 0.02 eV/atom are not shown.}
\end{figure}

The noble gas He becomes reactive at high pressure, leading to several new compounds, including Na$_2$He~\cite{dong2017stable}, HeN$_4$~\cite{li2018route}, He--alkali oxides (sulfides)~\cite{sun2014formation}, He--Fe~\cite{PhysRevLett.121.015301}, FeO$_2$He~\cite{PhysRevLett.121.255703}, Mg(Ca)F$_2$,~\cite{liu2018reactivity} He--H$_2$O~\cite{PhysRevB.91.014102,liu2019multiple}, and He--NH$_3$~\cite{bai2019electrostatic,PhysRevX.10.021007}. The incorporation of inert He tends to form open-framework structures with weak interactions between He and the host sublattice. For example, our previous calculations predicted a HeN$_4$~\cite{li2018route} compound formed at high pressure, which consists of open channels of N atoms holding He. Their weak interactions allow the removal of the He from the structure, leading to a pure $t$-N structure. Therefore, He may be regarded as a good intermediate for preparing new materials. The $t$-N phase obtained from high-pressure HeN$_4$~\cite{li2018route} motivated us to study whether new Si allotropes could be formed from high pressure Si--He compounds. A recent molecular dynamics (MD) simulation has demonstrated that Si and He react to form hP6-Si$_2$He at 7 GPa and 1500 K~\cite{bi2018formation}, which is a host-guest structure comprising a hexagonal diamond Si sublattice encapsulating He atoms.

\begin{figure}[htp]
\centering
  \includegraphics[width=0.9\linewidth,angle=0]{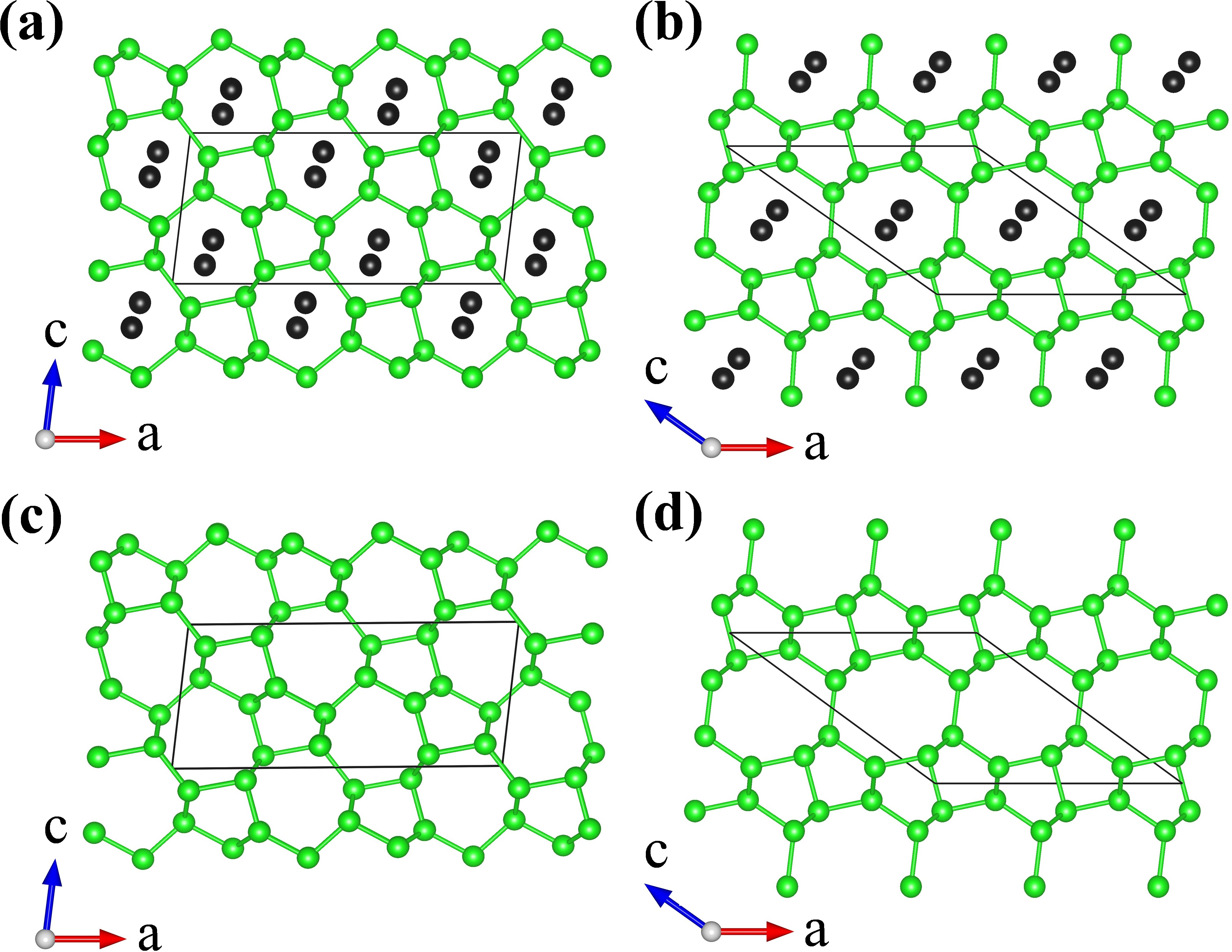}
  \caption{\label{fig:2} Crystal structures of \textbf {(a)} mC24-Si$_2$He, \textbf{(b)} mC16-Si$_3$He, \textbf{(c)} mC16-Si, and \textbf{(d)} mC12-Si. Black and green spheres represent He and Si atoms, respectively.}
\end{figure}

This work reports extensive structure searches on Si--He systems that predict two stable channel-like compounds (mC24-Si$_2$He and mC16-Si$_3$He) in addition to hP6-Si$_2$He~\cite{bi2018formation}. The He atoms trapped inside the channels are easily removed from mC24-Si$_2$He and mC16-Si$_3$He to form mC16-Si and mC12-Si, respectively. Interestingly, mC24-Si$_2$He, mC16-Si$_3$He, and mC12-Si are direct-gap semiconductors with band gaps of 1.13-1.34 eV. Importantly, mC24-Si$_2$He has an electric-dipole-transition allowed direct band gap, making it a good candidate photovoltaic material.

Structure predictions for the Si--He system were performed using CALYPSO~\cite{PhysRevB.82.094116,wang2012calypso}, which has correctly predicted many stable compounds under high pressure~\cite{zhu2014reactions,li2014metallization,PhysRevLett.115.105502,PhysRevLett.114.125501,PhysRevB.93.020103,cui2019role,liu2018effect,Xumeiling,CuiCSH7,shi2018nitrogen}. The structural optimization and electronic and optical properties were calculated using density functional theory as implemented in the Vienna $ab$ initio simulation package~\cite{PhysRevB.54.11169}, adopting the Perdew-Burke-Ernzerhof exchange-correlation functional under the generalized gradient approximation.~\cite{PhysRevB.46.6671,PhysRevLett.77.3865} The Heyd-Scuseria-Ernzerhof (HSE06) hybrid functional was employed to correct the electronic band structures.~\cite{heyd2003hybrid} All-electron projector augmented wave pseudopotentials with 1$s^2$ and 3$s^2$3$p^2$ valence configurations were chosen for He and Si atoms, respectively.~\cite{PhysRevB.59.1758} A plane wave cutoff energy of 800 eV and k-point mesh
of 2$\pi$ $\times$ 0.03 \AA$^{-1}$ were set to ensure total energy and forces convergence better than 1 meV/atom and 1 meV/\AA, respectively. Phonon calculations were carried out using a supercell approach as implemented in PHONOPY code.~\cite{PhysRevB.78.134106} First-principles MD simulations using $N$ (number of particles), $V$ (volume), and $T$ (temperature) were performed at 0 GPa and 300 K~\cite{nose1984unified}. 1 $\times$ 3 $\times$ 2 supercells for mC24-Si$_2$He (144 atoms) and mC16-Si (96 atoms), and 2~$\times$ 3~$\times$ 2 supercells for mC16-Si$_3$He (196 atoms) and mC12-Si (144 atoms) were employed. The migration barriers were calculated using the climbing image nudged elastic band (CI-NEB) method~\cite{henkelman2000climbing} based on supercells containing one He atom and 48 host Si atoms for both mC24-Si$_2$He and mC16-Si$_3$He. VASPKIT~\cite{vaspkit} was used to resolve the results of the transition dipole moment and the optical absorption spectra (the imaginary part of the dielectric function, $\varepsilon_2$).

\begin{figure}[htp]
\centering
  \includegraphics[width=0.85\linewidth,angle=0]{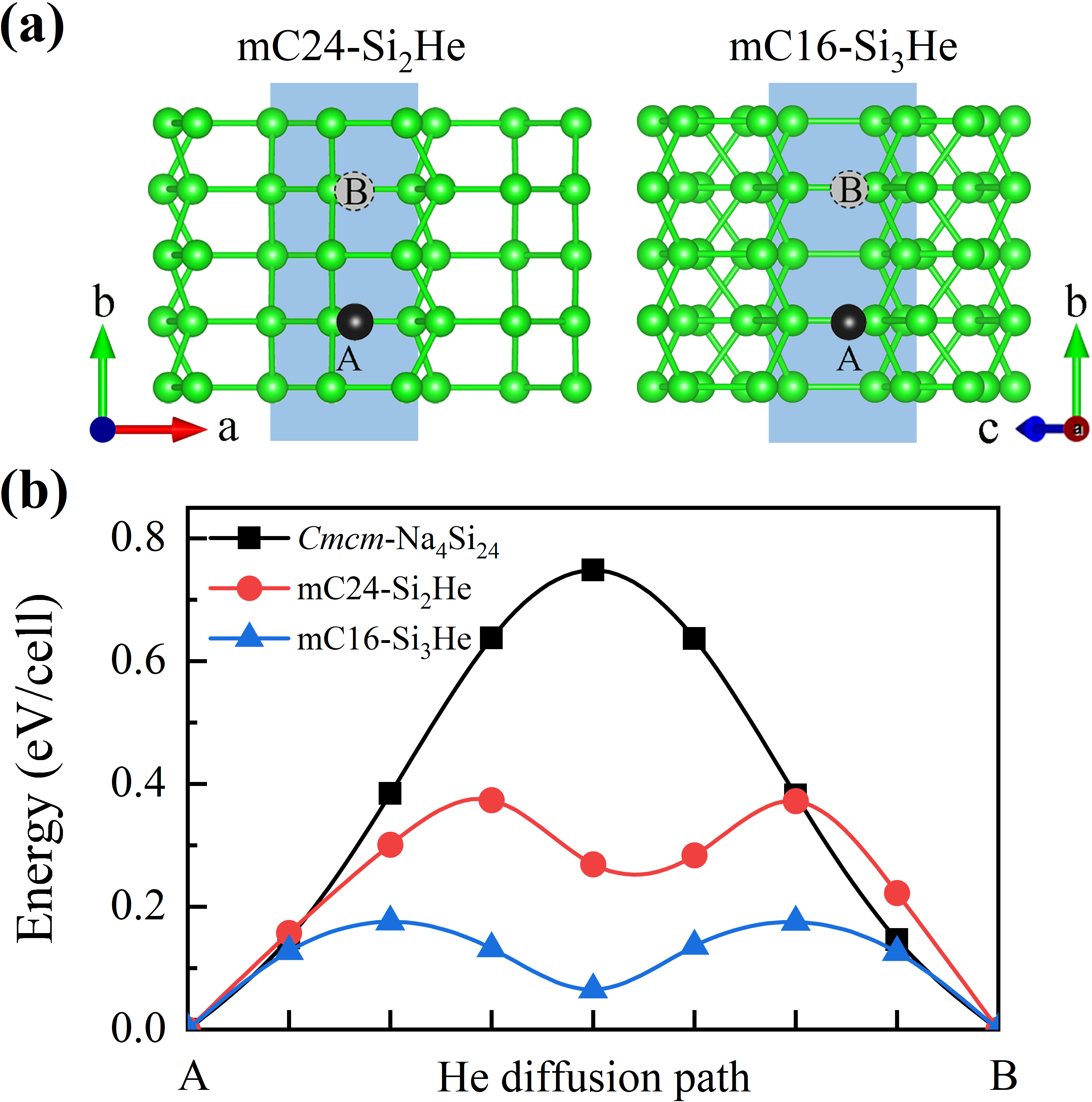}
  \caption{\label{fig:3} \textbf{(a)} Migration pathways of He atoms from site A to site B along the channels in mC24-Si$_2$He and mC16-Si$_3$He. The shaded regions indicate the longitudinal section of the channels. \textbf{(b)} Energy barriers for He migration along the channels at zero pressure, as well as Na migration in $Cmcm$-Na$_4$Si$_{24}$.~\cite{kim2015synthesis} }
\end{figure}

Structure predictions are first performed for Si$_2$He at 10 GPa with a maximum of eight formula units (f.u.) in a simulation cell. The previously proposed hP6-Si$_2$He~\cite{bi2018formation} is successfully predicted, but with much higher enthalpy, as shown in Fig.~\ref{fig:1}(a). Instead, the energetically most stable structure for Si$_2$He is mC24-Si$_2$He, which is monoclinic with space group $C2/m$ (8 f.u. in a unit cell) and is~$\sim$0.05 eV/f.u. energetically lower than hP6-Si$_2$He~\cite{bi2018formation}. Static-lattice enthalpy calculations reveal that mC24-Si$_2$He remains energetically most stable up to 17.4 GPa, above which hP6-Si$_2$He~\cite{bi2018formation} takes over, see Fig.~\ref{fig:2}(a). A previous MD simulation suggests that hP6-Si$_2$He~\cite{bi2018formation} could be formed at 7 GPa and 1500 K. Therefore, we examine the effect of temperature on the relative stability of the two structures using the quasi-harmonic approximation and find that temperature does not change the phase diagram of Si$_2$He, but rather postpones the transition pressure to 25 GPa at 1500 K. This result indicates that the newly predicted mC24-Si$_2$He phase is more favorable than hP6-Si$_2$He in experimental synthesis at low pressures.

In mC24-Si$_2$He, each Si atom connects to four other Si atoms to form three-dimensional networks with bond lengths of 2.43 \AA. Two kinds of channels sharing edges are found along the b-axis formed by five- or seven-membered rings of Si atoms. A zigzag arrangement of He atoms is located inside the larger channels formed by the seven-membered rings (see Supplemental Material~\cite{suppe}, Fig. S1). The shortest distance between He and the Si channel is 2.59 \AA, which is shorter than the Na--Si distance (3.01 \AA) in Na$_4$Si$_{24}$~\cite{kim2015synthesis}. Similar host-guest structures have been reported in several other compounds, such as Na$_4$Si$_{24}$~\cite{kim2015synthesis} and HeN$_4$~\cite{li2018route}. The previously proposed hP6-Si$_2$He~\cite{bi2018formation} can also be regarded as a host-guest structure with a distorted diamond hexagonal host Si lattice encapsulating guest He atoms inside the hexagonal channels. The lower enthalpy of mC24-Si$_2$He compared with hP6-Si$_2$He suggests that Si can form larger channels for the incorporation of He.

\begin{figure}[htp]
\centering
  \includegraphics[width=0.99\linewidth,angle=0]{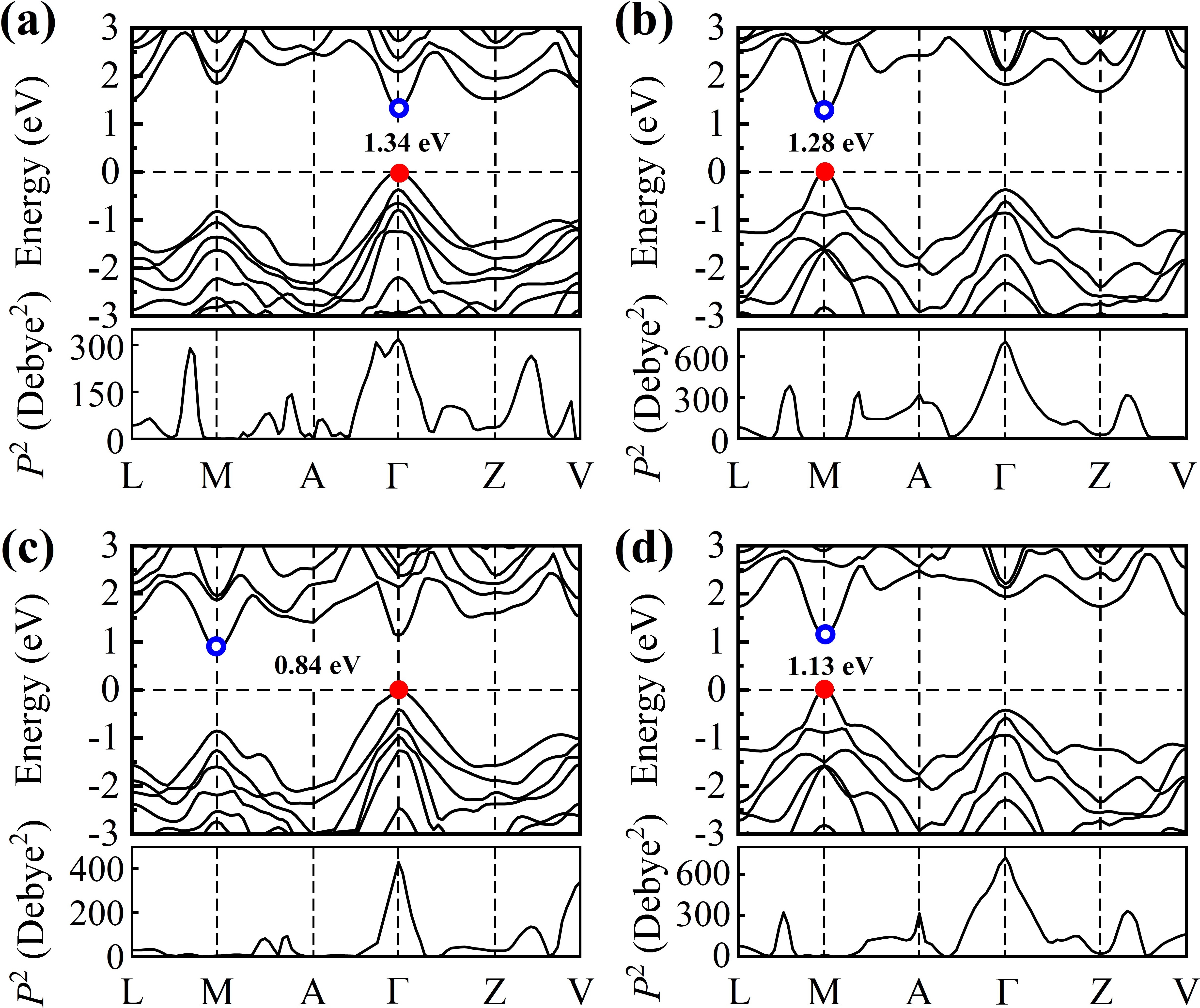}
  \caption{\label{fig:4} Electronic band structures of \textbf{(a)} mC24-Si$_2$He, \textbf{(b)} mC16-Si$_3$He, \textbf{(c)} mC16-Si and \textbf{(d)} mC12-Si at 0 GPa calculated based on the HSE06 functional. Red solid and blue hollow circles represent the valence band maximum and conduction band minimum, respectively. The lower panels in each figure are the square of the transition dipole moment.~\cite{meng2017parity}    }
\end{figure}

To search for other possible stable Si--He compounds, structural predictions are also performed for Si$_x$He$_y$ ($x$ = 1--12 and $y$ = 1--4) at 10 GPa using a maximum of 40 atoms in a simulation cell. Fig.~\ref{fig:2}(b) summarizes the formation enthalpies of the stoichiometries with respect to decomposition into mC24-Si$_2$He and CD-Si. Surprisingly, a new stable compound with stoichiometry Si$_3$He is identified with a negative formation enthalpy. The energetically most stable structure is mC16-Si$_3$He, which is monoclinic with space group $C2/m$ (4 f.u. in a unit cell), as shown in Fig.~\ref{fig:1}(b). mC16-Si$_3$He shares similar structural motifs with mC24-Si$_2$He, having a host-guest structure with Si-channels filled with He atoms. Nearly identical channels formed by five-membered rings of Si are observed in both mC16-Si$_3$He and mC24-Si$_2$He. mC16-Si$_3$He has a higher ratio of Si than mC24-Si$_2$He, which leads to larger channels formed by eight-membered Si rings enclosing He zigzag chains with a Si-He distance of 2.81 \AA. The dynamic stability of mC24-Si$_2$He and mC16-Si$_3$He at 10 and 0 GPa is confirmed by phonon dispersion calculations. The MD simulation reveals that both compounds exhibit thermodynamic stability at ambient pressure and temperature (300 K), suggesting that both could be quenched and recovered at ambient conditions once formed (see Figs. S2 and S3).

\begin{figure}[htp]
\centering
  \includegraphics[width=0.85\linewidth,angle=0]{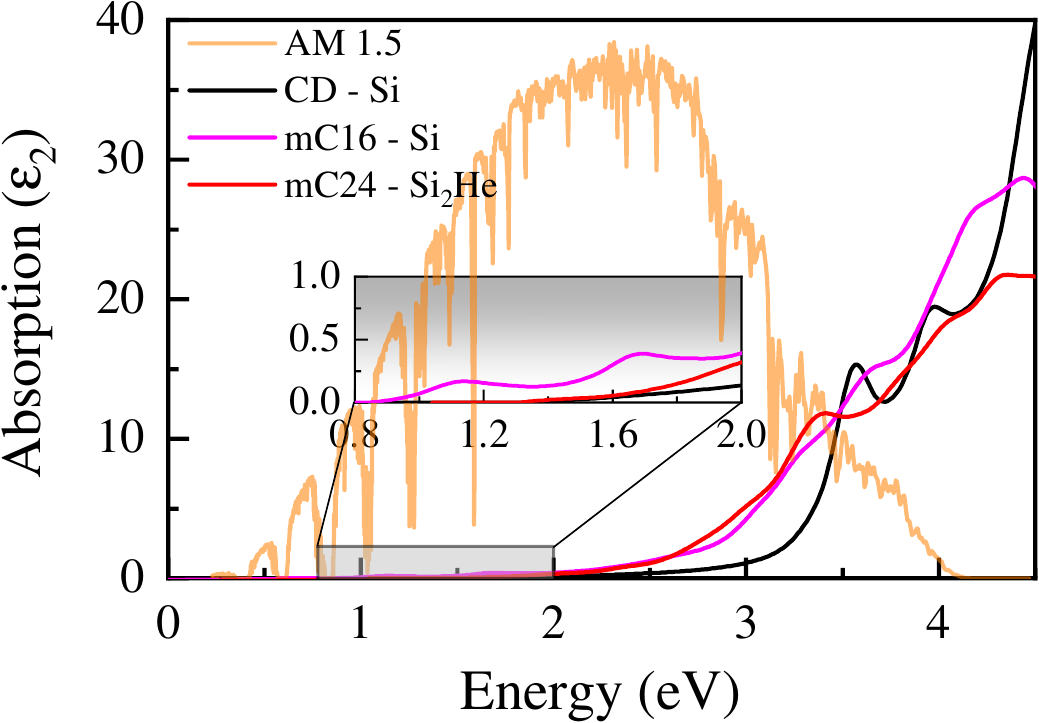}
  \caption{\label{fig:5} Imaginary part of the dielectric functions of various Si–He compounds and Si allotropes calculated with the HSE06 functional, as well as the reference air mass 1.5 (AM1.5) solar spectral irradiance~\cite{ASTM}. The inset shows a zoom in energy range of 0.8--2.0 eV for clarity.} 
\end{figure}

Electron localization function calculations exclude the existence of Si--He covalent bonds in both compounds given the absence of electron localization between them (see Fig. S4).~Bader charge analysis~\cite{henkelman2006fast} suggests slight charge transfer from the Si framework to each He atom of 0.05 electrons in mC24-Si$_2$He and 0.04 electrons in mC16-Si$_3$He, similar to those predicted in Na$_2$He~\cite{dong2017stable} and FeO$_2$He~\cite{PhysRevLett.121.255703}. The weak interaction between the Si frameworks and He atoms indicates the possible removal of He from the structures. Therefore, we examine the energy barriers of He diffusing along the channels, see Fig.~\ref{fig:3}(a). CI-NEB calculates energy barriers of 0.37 and 0.18 eV for mC24-Si$_2$He and mC16-Si$_3$He, respectively. These barriers are much lower than that (0.75 eV) faced when removing Na from Cmcm-Na$_4$Si$_{24}$~\cite{kim2015synthesis}, see Fig.~\ref{fig:3}(b), indicating comparatively easy removal of He atoms from mC24-Si$_2$He and mC16-Si$_3$He.

Figs.~\ref{fig:3}(c) and (d) show two pure Si structures obtained by removing He from mC24-Si$_2$He and mC16-Si$_3$He, denoted as mC16-Si and mC12-Si, respectively. Both Si allotropes retain Si frameworks nearly identical to those of the corresponding compounds. Phonon dispersion and MD calculations confirm the stability of both allotropes (see Figs. S2 and S3). A literature survey surprisingly found that the two Si structures have been previously predicted with much higher energies ($\sim$80 meV) than CD-Si~\cite{wang2014direct,wu2011density}. Metastable structures with higher energies are generally difficult to synthesize directly. Here, we provide a potential chemical pathway for the synthesis these two metastable Si allotropes, namely removing He atoms from pressure-stabilized Si–He compounds by thermal degassing.

Photovoltaic materials require a suitable direct band gap to ensure a large overlap with the solar spectrum in the visible range, and thus strong solar absorption. Electronic structures calculated on basis of the HSE06 functional reveal that both mC24-Si$_2$He and mC16-Si$_3$He have direct band gaps of 1.34 and 1.28 eV, respectively, close to the Shockley–Queisser limit (1.34 eV). Interestingly, after the removal of He atoms, mC12-Si retains a direct gap, although it is slightly decreased to 1.13 eV. In contrast, mC16-Si gains a quasi-direct gap with direct band gap of 1.12 eV at the $\Gamma$ point, which is slightly larger than the indirect band gap of 0.84 eV located between the $\Gamma$ and M points. The retained direct band gap in mC12-Si suggests a weak interaction between He and the Si channels in mC16-Si$_3$He, which is verified by the tiny charge transfer (0.04 eV) between them, as well as the negligible volume collapse (3\%) after removal of the He atoms. Compared with mC16-Si$_3$He, the removal of He from mC24-Si$_2$He distinctly changes the Si framework, which undergoes a 7\% volume collapse and local deformation (see Table S2), resulting in an indirect band gap in mC16-Si.

A direct band gap does not in itself guarantee good absorption, there should also be a dipole-allowed direct transition. Therefore, further calculation of the square of the transition dipole
moment ($P^2$) explores the transition permissibility between the direct band gaps. Interestingly, mC24-Si$_2$He shows a dipole-allowed direct transition with large $P^2$ value at the $\Gamma$ point, suggesting good potential as a photovoltaic material. The indirect band gap in mC16-Si is dipole-forbidden, but the large $P^2$ value related to the direct band gap at the $\Gamma$ point provides the possibility of good absorption. In contrast, mC16-Si$_3$He, mC12-Si, and the previously proposed hP6-Si$_2$He~\cite{bi2018formation} exhibit a dipole-forbidden direct band gap in view of the corresponding zero $P^2$ value, excluding them as good photovoltaic absorbers. Fig.~\ref{fig:5} compares the calculated imaginary parts of the dielectric constant of mC24-Si$_2$He, mC16-Si and CD-Si. Optical absorption in mC24-Si$_2$He starts at $\sim$1.3 eV, confirming that the direct band gap is dipole-allowed. It is evidenced that both mC24-Si$_2$He and mC16-Si have much better solar absorption capacities than CD-Si, as indicated by their broader overlap with the AM1.5 solar spectrum~\cite{ASTM}.

He, which has two electrons, is the most chemically inert natural element, although several recent works have predicted or synthesized He-containing compounds~\cite{li2018route,bi2018formation}. Despite this, He can be regarded as chemically inert in Si-He, as the atoms are almost completely independent of the surrounding structure with negligible charge gained from Si. Nonetheless, the current results provide evidence that the incorporation of He helps to stabilize new Si frameworks with weak van der Waals interactions. He appears to be chemically inert in all its known compounds (e.g. Na$_2$He~\cite{dong2017stable} and HeN$_4$~\cite{li2018route} ), allowing it to be removed easily from the surrounding structure without changing the structure substantially. Importantly, the removal of He hardly alters the charge distribution of the Si framework owing to the negligible charge transfer, allowing the electronic structures to be retained after He removal. This is confirmed by the direct band gap being retained in mC12-Si formed from mC16-Si$_3$He. Therefore, He appears to be a good intermediate for designing new functional materials.

In conclusion, extensive structure searches of Si--He systems predicted two dynamically stable compounds (mC24-Si$_2$He and mC16-Si$_3$He) with open framework structures comprising Si channels containing zigzag arrangements of He atoms. CI-NEB calculations revealed that the He could be easily removed along the channels in mC24-Si$_2$He and mC16-Si$_3$He to leave the pure Si allotropes, mC16-Si and mC12-Si, respectively. There were direct band gaps found in the electronic structures of mC24-Si$_2$He, mC16-Si$_3$He, and mC12-Si, whereas mC16-Si showed a quasi-direct band gap. The dipole-allowed direct band gap of 1.34 eV in mC24-Si$_2$He makes it a potential thin-film photovoltaic material. The current results demonstrate that He is an excellent element for regulating the properties of materials, as well as a good intermediate to synthesize functional materials.

The authors acknowledge funding from the NSFC under grants No. 11722433, No. 11804129, No. 11804128 and No. 11904142, and the funding from the Science and Technology Project of Xuzhou under grant No. KC19010. Y. L. acknowledges the funding from the Six Talent Peaks Project and 333 High-level Talents Project of Jiangsu Province. All the calculations were performed at the High Performance Computing Center of the School of Physics and Electronic Engineering of Jiangsu Normal University.

The authors declare no competing financial interests.



\end{document}